\documentclass[12pt]{article}

\newcommand{\eq}{\begin{equation}}
\newcommand{\eqx}{\end{equation}}
\newcommand{\eqn}{\begin{eqnarray}}
\newcommand{\eqnx}{\end{eqnarray}}
\newcommand{\f}[2]{\frac{#1}{#2}}

\newcommand{\tr}{\mbox{\rm tr}\,}

\newcommand{\Dl}{\Delta}
\newcommand{\al}{\alpha}
\newcommand{\bt}{\beta}
\newcommand{\dl}{\delta}
\newcommand{\lm}{\lambda}
\newcommand{\cor}[1]{\left\langle{#1}\right\rangle}
\newcommand{\brkt}[3]{\left\langle{#1}|{#2}|{#3}\right\rangle}
\newcommand{\qqqq}{\quad\quad\quad\quad}
\newcommand{\OO}{{\cal O}}
\newcommand{\TT}{{\cal T}}
\newcommand{\NN}{{\cal N}}
\newcommand{\ket}[1]{\left|{#1}\right\rangle}

\newcommand{\arr}[4]{%
\left(\begin{tabular}{cc}
$#1$ & $#2$ \\
$#3$ & $#4$
\end{tabular}\right)}

\newcommand{\hlc}{H^{l.c.}_{string}}
\newcommand{\opn}{\OO^J_{12,n}}
\newcommand{\opmr}{\TT^{J,r}_{12,m}}
\newcommand{\opr}{\TT^{J,r}_{12}}

\title{BMN operators and string field theory}

\author{Romuald A. Janik\footnote{
e-mail: {\tt janik@nbi.dk}}\\ \\
The Niels Bohr Institute,\\
Blegdamsvej 17, DK-2100 Copenhagen,\\ 
Denmark\\
and\\
Jagellonian University,\\
Reymonta 4, 30-059 Krakow,\\
Poland}

\begin{document}

\maketitle

\begin{abstract}
We extract from gauge theoretical calculations the matrix elements of
the SYM dilatation operator. By the BMN correspondence this should
coincide with the 3-string vertex of light cone string field theory in
the pp-wave background. We find a mild but important discrepancy with
the SFT results. If the modified $O(g_2)$ matrix elements are used,
the $O(g_2^2)$ anomalous dimensions are exactly reproduced without the
need for a contact interaction in the single string sector.
\end{abstract}

\section{Introduction}

In \cite{BMN} Berenstein, Maldacena and Nastase studied a pp-wave
limit of string theory in the $AdS_5 \times S^5$ background. Type IIB
strings on the pp-wave geometry were found to correspond to operators
of a $\NN=4$ SU(N) super Yang-Mills theory with large R charge $J$ in the
limit where $J^2/N$ is fixed. They obtained definite predictions for
the scaling dimensions of the relevant operators in the free string
limit which were subsequently verified on the gauge theory side
\cite{BMN,Gross1,Santambrogio}. 

Subsequent work was made in extending the correspondence on both sides
to lowest orders in the effective gauge coupling $\lm'=g^2_{YM}N/J^2$
and genus $g_2=J^2/N$ parameter \cite{K1,SEVEN,K2,CFHM,K3}. On the string
theory side the tool used to study interactions was light cone IIB
string field theory (SFT) constructed for the pp-wave background in
\cite{SV1} (and also inherently discrete string-bit formulations
\cite{Verl1,Zhou,Verl2}). There exist explicit expressions for the gauge
theory parameters $g_2$, $\lm'$ in terms of string theoretical
quantities (but see also \cite{Gopakumar}: 
\eq
\lm'=\f{1}{(\mu p^+ \al')^2} \qqqq g_2=4 \pi g_s (\mu p^+ \al')^2
\eqx
 
The link was made through a proposal made in \cite{SEVEN} of a relation between
matrix elements of the SFT hamiltonian and certain gauge theoretical
3-point functions. This was verified in various cases
\cite{Huang,Chu1,Kiem,Lee,SV2,Chu2} (see also
\cite{Klebanov,Gursoy,Russo,Gross2} for further developments). However 
the explicit proposal was not derived from `first principles'. A direct
calculation of the $O(g_2^2)$ anomalous dimensions from the $O(g_s)$
SFT matrix elements failed to give an agreement with the gauge
theoretical result. This was not a direct contradiction, however, due
to the theoretical possibility of $O(g_2^2)$ contact terms in the SFT
hamiltonian. In this letter we want to look for a more direct test of the SFT
-- gauge theory correspondence.

The main aim of this paper is to extract directly from the gauge theoretical
calculations done so far the order $O(g_2)$ matrix elements of the gauge theory
dilatation operator. This should be identified with the $O(g_s)$
vertex of light cone string field theory thus allowing for a direct
comparision with the formulation of \cite{SV1}. In addition it might give
some insight into the failure of SFT (modulo contact terms)
to describe the $O(g_2^2)$ gauge theoretical anomalous dimensions.  

The outline of this paper is as follows. In section 2 we will recall
some features of the BMN operator-string correspondence, in section 3 we will
extract the $O(g_2)$ matrix elements and show that they are sufficient
to reconstruct full $O(g_2^2)$ anomalous dimensions without the
need for explicit $O(g_2^2)$ `contact interactions' in the single
string sector. We conclude the paper with a discussion. 

\section{BMN operator-string correspondence}   

The dictionary established in \cite{BMN} between string theory and gauge
theoretical operators associates to each physical state of the string
an explicit (single-trace) operator of the gauge theory. The operators
which we will consider here are
\eqn
O^J &=& \f{1}{\sqrt{J N^J}} \tr Z^J \\
O^J_i &=& \f{1}{\sqrt{N^{J+1}}} \tr \left(\phi_i Z^J \right) \\
O^J_{ij,n}&=& \f{1}{\sqrt{J N^{J+2}}} \left( \sum_{p=0}^J e^{2\pi i p
n/J} \tr \left( \phi_i Z^p \phi_j Z^{J-p} \right) \right)
\eqnx
here $Z=(\phi_5+\phi_6)/\sqrt{2}$ and $\phi_i$ are other transverse
coordinates. These operators correspond respectively to the states
$\ket{0,p^+}$, ${a_0^i}^\dagger \ket{0,p^+}$ and ${a_n^i}^\dagger
{a_{-n}^j}^\dagger \ket{0,p^+}$. 


Double trace operators correspond to two-string states and at zero
genus ($g_2=0$) can be identified unambigously. The operators that we
will use here are
\eqn
\opr &=& \OO^{r J}_{12,n} \OO^{(1-r)J} \\
\opmr &=& \OO^{r J}_1 \OO^{(1-r) J}_2
\eqnx
Here $r\in (0,1)$ denotes the fraction of light cone momentum carried
by the first string. Presumably (bosonic) multistring states have to be
symmetrized (this will not be important here).

The light cone string hamiltonian is 
\eq
\hlc=\f{\mu}{2} \left(\Delta-J\right)  
\eqx
Therefore we should identify it (up to the factor $2/\mu$ and the
constant shift) as equivalent to the gauge theory dilatation
operator $D$. 

At zero-genus all the single and double string states are eigenstates
of $\hlc$ as the respective gauge theory operators are eigenstates of
$D$. Once we turn on the interaction, the dilatation operator will
start to mix the operators and $\hlc$ will start to mix the 
corresponding single and multi-string states. We expect the 
action of the full interacting operators $D$ and $\hlc$ 
on the gauge theory operators, and (multi-)string states respectively 
to coincide\footnote{Up to possible rescalings of the individual states.}:
\eqn
DO_\al &=& D_{\al\bt} O_\bt \\
\left(\f{2}{\mu} \hlc +J \right) \ket{\al} &=& h_{\al\bt} \ket{\bt}
\eqnx
i.e. we should have $D_{\al\bt}=h_{\al\bt}$. 

In \cite{SV1} the terms linear in $g_s$ in $\hlc$ were constructed:
\eq
\hlc=H^{l.c.}_2+g_s H^{l.c.}_3
\eqx
where $H^{l.c.}_2$ is the free hamiltonian and $H^{l.c.}_3$ represents
the 3-string vertex. The following matrix elements were computed in
\cite{SV2} and will be relevant later:
\eqn
\label{e.svel1}
\brkt{\opn}{H^{l.c.}_3}{\opmr}&=& \f{4 \mu}{\pi} (1-r) \f{nr+m}{nr-m}
\sin^2(\pi n r) \\
\label{e.svel2}
\brkt{\opn}{H^{l.c.}_3}{\opr}&=& \f{4 \mu}{\pi} \sqrt{r(1-r)} \sin^2(\pi n r)
\eqnx

Up till now most comparisions between string field theory and gauge
theory were performed either on the level of 3-point correlation
functions or by computing
scaling dimensions. 

The former method was based on a proposal which
linked the structure constants $C_{ijk}$ and appropriate SFT hamiltonian matrix
elements \cite{SEVEN}
\eq
\label{e.proposal}
\brkt{i}{H^{l.c.}_3}{j,k}=\mu g_2(\Dl_i-\Dl_j-\Dl_k) C_{ijk}
\eqx
Although plausible and supported by various calculations it has not
been strictly proven from first principles nor
shown how it could be systematically extended beyond leading order. 

The latter method of comparison based on determining anomalous
dimensions is difficult because 
the first nontrivial corrections to the scaling dimensions are of order
$O(g_2^2)$ while the SFT hamiltonian in the pp-wave background has been
only determined to $O(g_2)$ order. Indeed $H^{l.c.}_3$ with the matrix
elements (\ref{e.svel1})-(\ref{e.svel2}) could not reproduce
\cite{SEVEN,CFHM} the $g_2^2$ correction to the anomalous dimension of
the $O^J_{ij,n}$ operator obtained in a SYM calculation \cite{K2}:
\eq
\label{e.exact}
\f{g_2^2 \lm'}{4 \pi^2} \left(\f{1}{12}+\f{35}{32 \pi^2 n^2} \right)
\eqx

In fact the disagreement between the
scaling dimensions calculated in gauge theory and ones obtained from
the cubic interaction hamiltonian has been attributed to the possible
appearance of nontrivial contact terms of order $O(g_s^2)$. Indeed
additional $O(g_s^2)$ terms appear also in flat space light cone SFT
\cite{GK1,GK2}.  However there they only involve four string fields
while here it seems that the disagreement can be cured only by terms
which involve only {\em two} string fields. 

Therefore it is interesting to directly extract the
$O(g_2)$ matrix elements of the gauge theory dilatation operator as
these, according to the BMN operator-string correspondence, should be
identified with the $O(g_s)$ SFT hamiltonian matrix elements.

\section{Gauge theory results}

We will now extract the matrix of the gauge theory dilatation operator
up to order $O(g_2)$. Let $O_\al$ be the set of all operators (single-
and multi-trace) with R charge $J$ which are eigenstates of the free
(planar) dilatation operator, $\bar{O}_\al$ the corresponding
complex conjugates and let us denote by $O'_A$ the operators with
definite scaling dimension:
\eq
D O'_A=\Dl_A O'_A
\eqx
where $D$ is the dilatation operator.
These $O'_A$'s may be rewritten as linear combination of the original operators
and vice-versa
\eq
\label{e.expan}
O_\al= V_{\al A} O'_A \quad\quad (O=VO') \qqqq O'=V^{-1}O
\eqx
Similar formulas hold for the barred operators (with a different
matrix\footnote{We do not need to assume anything about the relation
of $V^*$ to $V$.} $V^*_{\al A}$).
Thus the matrix elements of the gauge theory dilatation operator in
the original basis $O_\al$ are
\eq
DO_\al \equiv D_{\al\bt}O_\bt = (V\Dl V^{-1})_{\al\bt} O_\bt
\eqx
This should be identified with $\f{2}{\mu}\brkt{\al}{\hlc}{\bt}
+J\dl_{\al\bt}$. 

We will now show how to extract the matrix $V\Dl V^{-1}$ from 2-point
correlation functions. Using the expansions (\ref{e.expan}) we get
\eq
\cor{O_\al(0) \bar{O}_{\bt}(x)}=V_{\al A} V^*_{\bt B} \f{\dl_{AB}
C_A}{|x|^{2(J+2+\Dl_A)}} 
\eqx
Here the $C_A$'s are some undetermined normalization constants.
Expanding to linear order in the logarithm gives
\eq
\cor{O_\al(0) \bar{O}_{\bt}(x)}=\f{1}{|x|^{2(J+2)}} \left(
M'_{\al\bt}+M''_{\al\bt} \log(x\Lambda)^{-2} \right)
\eqx
where the matrices $M'$ and $M''$ are given by
\eq
M'=VCV^\dagger \qqqq M''=VC\Dl V^\dagger
\eqx
and $V^\dagger$ denotes here the transpose of $V^*$. 
The dilatation operator matrix is then given by
\eq
D_{\al\bt}=(M'' {M'}^{-1})_{\al\bt}
\eqx

The matrices $M'$ and $M''$ have been calculated in \cite{K2}. For our
purposes it is enough to find their elements to order $O(g_2)$. To
this order there are 
only nonzero elements in the $\opn$ -- $\opmr$ sector and the $\opn$ --
$\opr$ sector. It is easy to see that to order $O(g_2)$ we may treat them
independently. 

\subsubsection*{The $\opn$ -- $\opmr$ sector}

The calculations of \cite{K2,CFHM}  yield (see e.g. (3.15) in
\cite{K2})
\eqn
M'&=&\arr{1}{g_2 x}{g_2 x}{1} \nonumber\\
M''&=&\lm' \arr{n^2}{g_2 x \left(\f{m^2}{r^2}+n^2-\f{mn}{r}\right)}{%
g_2 x \left(\f{m^2}{r^2}+n^2-\f{mn}{r}\right)}{\f{m^2}{r^2}}
\eqnx
with
\eq
x=\f{r^{3/2} \sqrt{1-r} \sin^2(\pi n r)}{\sqrt{J} \pi^2 (m-n r)^2}
\eqx
The dilatation matrix to order $O(g_2)$ is thus
\eq
\label{e.vertnmr}
\arr{\lm' n^2}{0}{0}{\lm' \f{m^2}{r^2}}+
g_2
\arr{0}{\lm'  x \f{m}{r^2}(m-nr)}{%
-\lm'  x \f{nr}{r^2} (m-nr)}{0}
\eqx
Several comments are in order here. Firstly the result does not agree
with the matrix elements of \cite{SV2}. There is some relation,
however. We note that the difference of the off-diagonal elements is
equal to
\eq
\f{4 g_s}{\pi \sqrt{J}} \sqrt{\f{1-r}{r}} \f{nr+m}{nr-m} \sin^2(\pi n r) 
\eqx
which exactly coincides with (\ref{e.svel1}) up to the normalization
factor of $\sqrt{Jr(1-r)}$.
In fact we see that the rhs of the proposal (\ref{e.proposal}) is
antisymmetric w.r.t exchange of initial and final states. A minor
generalization which would still hold even for the modified matrix
elements (\ref{e.vertnmr}) would be 
\eq
\f{1}{2}\left( \brkt{i}{H^{l.c.}_3}{j,k} - \brkt{j,k}{H^{l.c.}_3}{i}
\right) = \mu g_2(\Dl_i-\Dl_j-\Dl_k) C_{ijk}
\eqx 
Secondly the matrix (\ref{e.vertnmr}) does not have a definite
symmetry. From the SFT 
point of view this would signify that the amplitude of splitting
strings is different from joining. This does not necessarily mean that
the gauge theory dilatation operator is non-hermitian since the
natural scalar product is non-zero only between the barred and
non-barred sectors. We will return to this point in the discussion.

\subsubsection*{The $\opn$ -- $\opr$ sector}

In this case the relevant formulas (see e.g. (3.15) in
\cite{K2}) are
\eq
M'=\arr{1}{g_2 y}{g_2 y}{1} \qqqq 
M''=\lm' \arr{n^2}{g_2 y n^2}{g_2 y n^2}{0}
\eqx
with 
\eq
y=\f{1}{\sqrt{J}}\left(\dl_{n,0} r-\f{\sin^2(\pi n r)}{\pi^2 n^2}
\right)
\eqx
The dilatation matrix to order $O(g_2)$ is thus
\eq
\label{e.vertnr}
\arr{\lm' n^2}{0}{0}{0}+
g_2
\arr{0}{0}{\lm' y n^2}{0}
\eqx
Again we see that it is nonsymmetric and that only the difference of
off-diagonal elements gives the SV matrix element (\ref{e.svel2}).

Let us now assume that the cubic $O(g_s)$ SFT vertex is given
by the above formulas (\ref{e.vertnmr}) and (\ref{e.vertnr}).
We will show that this is enough to reproduce the exact
gauge-theoretic scaling dimension to order $O(g_2^2)$.

\subsection*{Scaling dimensions to order $O(g_2^2)$}

The formulas for scaling dimension follow easily (as in
\cite{SEVEN,CFHM}) from first 
order perturbation theory in the off-diagonal elements of the
hamiltonian (dilatation matrix), but keeping in mind the fact that the
hamiltonian is non-symmetric.
Indeed assuming that $D_{\al\bt}=\Delta_\al \dl_{\al\bt} +g_2
H^{(1)}_{\al\bt}+g_2^2 H^{(2)}_{\al\bt}$ we obtain
\eq
\Delta=\Delta_\al +g_2^2 \sum_\bt \f{H^{(1)}_{\al\bt}
H^{(1)}_{\bt\al}}{\Delta_\al-\Dl_\bt} +g_2^2 H^{(2)}_{\al\al} 
\eqx
We assume that $H^{(2)}_{\al\al}=0$ (no contact interactions in the
single string sector).
We will now show that the full $O(g_2^2)$ result is obtained. It is
interesting to compare with section 5.2 in \cite{SEVEN}. Now $\opr$
does {\em not} contribute as the product of the off-diagonal elements in
(\ref{e.vertnr}) vanishes. Only the operators $\opmr$ give a
contribution. Since $\Delta_n-\Delta^r_m=\lm'(n^2-m^2/r^2)$ we have
to calculate
\eq
g_2^2 \sum_{m,r} \f{-\lm' x^2 \f{nm}{r^3} (m-nr)^2}{n^2-\f{m^2}{r^2}}=
-\f{g_2^2 \lm'}{J \pi^4}\sum_{m,r} r^2(1-r) \sin^4(\pi n r) \f{n m}{(m-nr)^2
(n^2 r^2-m^2)}
\eqx
We now use the formula
\eqn
\!\!\!\!\!\sum_{m=-\infty}^\infty \f{n m}{(m-nr)^2 (n^2 r^2-m^2)}\!\! &=& \!\!
\f{\pi}{4 n r^2} \bigg( -n\pi r \csc^2(n\pi r) + \nonumber\\
&&\!\! + \cot(n \pi r) ( 2n^2 \pi^2 r^2 \csc^2(n \pi r)-1) \bigg)
\eqnx
and replace $(1/J)\sum_r$ by an integral. The result is
\eq
\label{e.result}
\f{g_2^2 \lm'}{4 \pi^2} \left(\f{1}{12}+\f{35}{32 \pi^2 n^2} \right)
\eqx
in agreement with (\ref{e.exact}).
We see that the full $O(g_2^2)$ result was obtained just from the cubic
$O(g_2)$ interaction. The positive sign of the correction for $n=1$
could only appear due to the fact that the matrix (\ref{e.vertnmr}) is
nonsymmetric. 
In comparision to the work of \cite{K2} the above result (\ref{e.result})
was derived here only from a small subset of data.
This is a strong argument in favour of a SFT
interpretation --- $O(g_2^2)$ elementary interactions (contact terms) in
the single string sector, which seem unlikely by comparision to the
flat space SFT indeed do not appear here (by the above calculation we
demonstrated that $H^{(2)}_{nn}=0$). 
On the gauge theory side, if
it were not for the SFT interpretation we would not have any reason to
expect a vanishing $O(g_2^2)$ term in the single trace (single string)
sector. 

However the main problem which remains is how to reconcile the
asymmetric SFT vertex reconstructed here from the gauge theory
calculations of \cite{K2} with the construction of light cone SFT in the
pp-wave background. 

\section{Discussion}

In this paper we have reconstructed the order $O(g_2)$ matrix elements
of the dilatation operator directly from gauge theory calculations. By
the BMN operator-string correspondence this should give the 3-string
$O(g_s)$ vertex of light cone SFT in the pp-wave background. We find a
disagreement with the continuum SFT matrix elements
of \cite{SV2} even at order $O(g_s)$.  

From this point of view we may return to the problem of the failure of
SFT to reproduce the correct gauge theory scaling dimensions.
Previously this was attributed to the possible existence of $O(g_s^2)$
contact terms. However from the flat space perspective such contact
terms in the {\em single} string sector are unlikely.

Here we show that there is a disagreement even at order
$O(g_s)$, although a mild one. With the `new' $O(g_s)$ matrix elements 
the full $O(g_2^2)$ anomalous dimensions can be reconstructed 
without any additional $O(g_2^2)$ contact terms. As was mentioned
earlier we believe that this is an argument in favour of a SFT
interpretation.  

The deviation from the matrix elements of the SFT vertex constructed
in \cite{SV1} is not very large. The asymmetric component coincides with
the SFT matrix elements of \cite{SV2}. So perhaps there is room for
reconciling these results with SFT.

A curious feature of the gauge theoretical dilatation matrix which we
obtained is that it does not have any simple symmetry
properties. Matrix elements which would correspond on the string
theory side to `splitting' and `joining' of strings are different.
From the point of view of string theory this asymmetry may not be
unacceptable as, in contrast to flat space, the pp-wave background is
not symmetric w.r.t light cone time reversal ($x^+ \to -x^+$) since
then the RR field strength changes sign.
On the gauge theory side there is no obvious contradiction with
hermiticity because the natural scalar product is off-diagonal and is
non vanishing only for operators with {\em opposite} R charge. It
would be interesting to see how it is possible to understand explicitly
that lack of symmetry within the SFT framework.     

A remaining open problem is to reproduce the dilatation matrix
elements derived here from `continuum' SFT. As this paper was being written
\cite{Verl2} appeared which gave a refined discrete string bit approach to the
BMN-string correspondence. It would also be interesting to examine the
interrelation with the framework of \cite{Gross2}.

\bigskip

\noindent{}{\bf Acknowledgments.} I would like to thank Charlotte
Kristjansen, Niels Obers, Jens Lyng Petersen, Shigeki Sugimoto and
Paulo di Vecchia for discussion. This work was supported by the EU
network on ``Discrete Random Geometry'' and KBN grant~2P03B09622.

\end{document}